
\magnification=1200
\hyphenpenalty=2000
\tolerance=10000
\hsize 14.5truecm
\hoffset 1.truecm
\openup 5pt
\baselineskip=24truept
\font\titl=cmbx12



\def\ref{\par\noindent\hangindent 20pt}
\def\refig{\par\noindent\hangindent 15pt}
\def\mincir{\raise
-2.truept\hbox{\rlap{\hbox{$\sim$}}\raise5.truept
\hbox{$<$}\ }}
\def\magcir{\raise
-4.truept\hbox{\rlap{\hbox{$\sim$}}\raise5.truept
\hbox{$>$}\ }}
\def\rho{\varrho}
\def\Mdot{\dot M}
\def\MS{M_*}

\def\rs{R_*}
\null
\vskip 1.truecm
\centerline{\titl OLD ISOLATED ACCRETING NEUTRON STARS:}
\smallskip
\centerline{\titl CONTRIBUTION TO THE SOFT X--RAY BACKGROUND}
\smallskip
\centerline{\titl IN THE 0.5--2 KEV BAND }
\vskip 1.truecm
\centerline{Silvia Zane $^1$, Roberto Turolla $^2$, Luca Zampieri $^1$,
Monica Colpi $^3$ and Aldo Treves $^1$}
\bigskip
\centerline{$^1$ International School for Advanced Studies, Trieste}
\centerline{Via Beirut 2--4, 34014 Trieste, Italy}
\medskip
\centerline{$^2$ Department of Physics, University of Padova}
\centerline{Via Marzolo 8, 35131 Padova, Italy}
\medskip
\centerline{$^3$ Department of Physics, University of Milano}
\centerline{Via Celoria 16, 20133 Milano, Italy}
\vfill
\bigskip\bigskip

\beginsection ABSTRACT

The issue of the observability of Old Isolated Neutron Stars (ONSs)
accreting from the interstellar gas, is reconsidered using the
spectra presented in Zampieri {\it et al.\/} (1995). In
particular,
we focus our attention on the overall soft X--ray emission of ONSs,
which may
provide a substantial contribution to the X--ray background in the 0.5--2
keV band. It is found that if accretion is channeled into polar caps by a
magnetic field of $\sim 10^9$ G, up to $\sim$
$12-25 \%$ of the unresolved soft excess observed at high latitudes in the
X--ray background can be explained in
terms of emission from accreting isolated ONSs. However, this would imply
the detection with ROSAT of $\sim 10$ sources deg$^{-2}$ above a threshold
of $\sim 10^{-14} \, {\rm erg\, cm^{-2} s^{-1}}$.
A brief reconsideration of the observability of isolated ONSs as individual
X--ray sources is also presented.

\bigskip\bigskip
\noindent
{\it Subject headings:\/} accretion, accretion disks \
-- \ background radiations \ -- \ stars: neutron \ -- \ X--rays: sources \ -- \
X--rays: stars\hfill\break
\vfill\eject

\beginsection 1. INTRODUCTION

Old neutron stars (ONSs), i.e. neutron stars which evolved beyond the
pulsar phase, may  be as many as $\sim 10^9$ in the Galaxy.
Even if their number accounts for a few percent of the total Galactic
star population, isolated ONSs have not been unambiguously detected so far.
Although the thermal emission resulting from
cooling of the hot interior is too weak for being observed
after a time $\sim 10$ Gyr, old isolated
neutron stars fed by the interstellar gas may show up as very weak,
soft X--ray sources, as originally suggested by Ostriker, Rees and Silk
(1970).
For a speed relative to the interstellar medium (ISM) of 10 ${\rm km \,
s}^{-1}$
and a density of 1 ${\rm cm}^{-3}$, the accretion luminosity $L$ of an
isolated ONS is $\sim 10^{31} \, {\rm erg \, s}^{-1}$.
If ONSs emit from their entire surface (of radius $\rs$) as a blackbody,
the effective temperature
$$
T_{eff} = \left( { {L}\over {4 \pi \rs^{2} \sigma}} \right)^{1/4}
$$
turns out to be $\sim 30$ eV.  A harder spectrum
would be emitted if ONSs retain, after $\sim 10$ Gyr,
a residual magnetic field which funnels the accretion flow in the polar cap
regions: for a field of $10^9$G, $T_{eff}$ could be 3 times larger
at the same accretion luminosity.
The low bolometric luminosity and the softness of the spectrum explain the
difficulty of observing an isolated accreting ONS.

The detection of ONSs was included as a possible target for
the {\it Einstein\/} mission (Helfand {\it et al.\/} 1980), but it was only a
decade later that
this issue came to life again when Treves and Colpi (1991, hereafter TC)
reconsidered the
observability of ONSs with ROSAT. Assuming a blackbody spectrum,
using the velocity distribution
proposed by Paczy\'nski (1990) and a local density $0.07 \ {\rm cm}^{-3}$ for
the ISM, they found that thousands of ONSs should appear in the
ROSAT PSPC All Sky Survey in the most favorable case of polar accretion.
A complete analysis by Blaes and Madau (1993, hereafter BM) essentially
confirmed TC results.

Recent observations with the ROSAT Wide Field Camera during survey phases
(Pounds {\it et al.\/} 1993) and with EUVE
have provided evidence for the presence of a relatively small number
($\sim 23$) of unidentified bright sources:
this led Madau \& Blaes (1994) to rule out there being as many
as $10^9$ isolated ONS in the Galaxy, if accretion is spherical.
Uncertainties on the statistical properties of the old neutron star
population and on the physics of the accretion process
make nonetheless
the issue of the total number of Galactic ONSs still controversial
(see e.g. Treves {\it et al.} 1995). The search for favorable sites
where accretion can make a lone neutron star shining is therefore of
importance. In connection with this problem, the detectability
of ONSs in molecular cloud
was examined by BM and Colpi, Campana, \& Treves (1993, hereafter CCT).
These studies were recently corroborated
by the work of Stocke {\it et al.} (1995), in which it is
proposed that a still unidentified source in the Einstein Extended Medium
Sensitivity Survey is an ONS in the Cirrus cloud. A systematic search of
isolated ONSs using ROSAT observations is presently under way (Danner, Kulkarni
\& Hasinger 1994).

The interest for the X--ray emission from isolated ONSs is not restricted to
the detection of individual sources. Despite their intrinsic weakness, in
fact, they could provide a contribution to the diffuse X--ray
background (XRB), being their total number very large. This point was already
discussed by BM, who found, however, that the integrated flux from ONSs
is much lower than the XRB above 0.5 keV.
In a recent paper Hasinger {\it et al.\/} (1993) have analyzed the
X--ray emission in selected high--latitude fields and found that the number of
resolved sources exceeds by $\sim $ 60\% the number of QSOs expected from a
standard evolutionary scenario. This led the suggestion
that a new population of sources can contribute sizably
to the soft X--ray background detected at 0.5--2 keV with ROSAT. Although
such a population may be extragalactic in origin, Maoz \& Grindlay (1995,
hereafter MG) showed that its properties are compatible with those of
galactic objects and tentatively identified them with Cataclysmic Variables
(CVs). Accreting ONSs
could have the desired spatial distribution but were ruled out
as possible candidates by MG on the basis of their too soft emission. We
stress, however, that both BM and MG restricted their discussion to the
very simple case in which the spectrum is a blackbody at $T_{eff}$,
emitted from the entire star surface.

In the light of the previous discussion, the issue of the correct
determination of the spectrum of an accreting ONS seems indeed to be
compelling.
In a recent paper (Zampieri, Turolla, Zane \& Treves 1995), we addressed this
problem presenting a self--consistent calculation of
the X--ray spectrum  emitted by a unmagnetized neutron star, solving the
full radiative transfer problem for a static, plane--parallel atmosphere.
It was found that, at very low accretion rates, corresponding to $L\mincir
10^{31} \ {\rm erg\, s}^{-1}$, emitted spectra show an overall hardening
with respect to the blackbody at $T_{eff}$,
even if the atmosphere is in LTE
and the scattering optical depth is lower than the absorption one;
typical hardening
factors ($\langle $ radiation temperature $\rangle$/$T_{eff}$)
are about 2.5. At energies larger than 0.5 keV, where the
blackbody emission becomes vanishingly small, the computed spectrum
shows a persistent tail and a fraction of 10\%--20\% of the total flux
is emitted above this threshold.

In this paper we reconsider the observational impact of accreting ONSs
using the calculated spectra and allowing for
polar cap accretion.
The paper is structured as follows: in section 2 we overview the main
points about the physics of the accretion process and in section 3 the present
distribution function of ONSs is calculated.
Section 4 contains our analysis of the contribution of ONSs to the XRB and
in section 5 estimates of the observability of individual sources are
presented.
Discussion and conclusions follow.

\beginsection 2. ACCRETION ONTO ISOLATED NEUTRON STARS

The total luminosity $L$ emitted in the accretion process
depends on the star velocity $v$ and
the density of the surrounding gas, $n$, and can be written as
(see e.g. Novikov \& Thorne 1973)
$$
\ell  = (1 - y){\dot m}
= 1.85 \times 10^{-3}(1 - y)n v^{-3} \eqno(1)
$$
\noindent where $n \, = \, (1 + \chi)n_H $, $\chi \, = \, 0.36$
for standard
chemical composition, $n_H$ is
the hydrogen number density, $\ell$, ${\dot m}$
are the luminosity and the accretion rate in
Eddington units, and the velocity is in km/s.
The factor $1 - y$, where
$y = \sqrt {(1 - R_g/\rs)}$ and $R_g$ is the gravitational radius, is the
relativistic efficiency;
typical values for the neutron
star mass and radius are used throughout: $M = 1.4 \MS$ and
$\rs = 3 \, R_g=12.6$ km. In the previous formula the star velocity is
replaced by the local sound speed of the interstellar gas $c_s$ when $v< c_s$.
Alternatively one could assume that $\ell\propto (v^2+c_s^2)^{-3/2}$. It
should be noted, however,
that when $v\sim c_s$, both approaches give an approximated value, no simple
expression for $\dot m$ being available (see again Novikov \& Thorne 1973).

In order to evaluate the accretion luminosity a value for $n_H$ must be
supplied.
The spatial distribution of the interstellar medium in the galaxy
is highly inhomogeneous (see e.g. Dickey \& Lockman (1990) for a detailed
review on this subject). Observational data from Ly$\alpha$ and 21 cm
absorption measures shows that the ISM distribution of both cold and
warm HI is nearly constant in radius while its $z$--dependence
can be fitted by
$$
n_{HI} = n_1 \exp {\left( -{ {z^2} \over { 2 \sigma_1^2 } } \right)} +
n_2 \exp{\left( -{ {z^2} \over { 2 \sigma_2^2 } } \right)} +
n_3 \exp{\left( -{z \over h }\right)} \eqno(2)
$$
with $n_1 = 0.395$, $n_2 = 0.107$, $n_3 = 0.064 $;
$\sigma_1 = 212$, $\sigma_2 = 530$ and $h = 403$ pc; $n_i$ and $\sigma_i$ are
in cm$^{-3}$ and pc, respectively. The applicability of the previous expression
is restricted to the range
$0.4 \leq R/R_0 \leq 1$, where $R$ is the galactocentric radius and
$R_0 = 8.5 $ kpc is the distance of the Sun from the Galactic center.
The gas layer has a scale height of about 230 pc in the vicinity of the Sun,
while
for $R \leq 0.4R_0$ it shrinks to $\approx 100$ pc and in the outer Galaxy it
expands linearly up to $\approx 2$ kpc.
The other important contribution to the total ISM density comes from
molecular hydrogen. The best tracer of H$_2$ is the CO molecule,
and observational data suggest a local gaussian distribution with a
scale height of $\sim 60-75$ pc. Observations, however, are much less
conclusive
as far as the midplane density is concerned (Bloemen 1987), also because it
may significantly depend on $R$ (De Boer 1991).
Here we approximate the H$_2$ distribution
with a gaussian with central density 0.6 ${\rm cm ^{-3}}$ and
FWHM 70 pc (De Boer 1991).
The ionized component gives only a very small contribution to the total
density, and it will be neglected.

The remaining free parameter in our discussion is the value of $c_s$. If
the star is moving supersonically through the ISM, the main contribution to
the accretion rate comes from the material swept by its motion.
On the other hand, we are mainly interested in the low--velocity tail of
the ONSs distribution because these are the stars which radiate the higher
luminosities. Their velocity may become lower than the
ISM sound speed, and the accretion rate is then fixed by the thermal velocity.
Since the accretion radius turns out to be always smaller than the Str\"omgren
radius by a factor $10^3$, $c_s$ is the sound speed in a fully ionized medium
at the equilibrium ionization temperature ($T \sim 10^4$ K)

$$ c_s\sim 10\sqrt{(T/10^4 \, {\rm K})} \ {\rm km \, s^{-1}}\sim 10 \
{\rm km\, s^{-1}}\, . \eqno(3)$$

Once $n_H$ and $c_s$ are fixed, the observability of an ONS
depends on its distance, velocity and on the response of the detector.
The count rate $N$, corrected for
the absorption due to the interstellar gas and for the detector
effective area $A_\nu $, is
$$
N = \int_{0}^{\infty} {F_\nu \over {h \nu}} \exp \left({-\sigma_{\nu} N_{H}}
\right) A_\nu d \nu \eqno(4)
$$
where $F_\nu=L_\nu/4\pi d^2$, $d$ is the distance of the source, $N_{H}$ is
the column density
and $\sigma_{\nu}$ is the absorption cross section (Morrison \& McCammon 1983).

Spectral calculations are presented in Zampieri {\it et al.\/} (1995) for
emission coming from the entire star surface and for $10^{-7}\leq \ell\leq
10^{-2}$. The lower limit corresponds to $L\simeq 2\times 10^{31}$ erg/s
for $\MS = 1.4 M_\odot$ and a pure H atmosphere. Since luminosities $\sim
10^{30}$ erg/s are expected from ONSs with velocities $\sim 20-30$ km/s, we
have calculated model spectra down to $\ell = 4\times 10^{-8}$; numerical
problems prevented us to reach lower luminosities. Spectra corresponding to
$\ell = 10^{-7},\, 4\times 10^{-8}$ are shown in figure 1.
A comparison between the actual emerging flux and the blackbody one
above 0.1 keV shows that the former is greater by 10\%--40\%, and that it
deviates more and more from a Planckian as luminosity decreases,
showing a broad maximum. In the following
sections we will be interested to evaluate the source luminosity in the
0.5--2 keV band, $L_x$, for any given star velocity $v$, or, equivalently,
for any given value of the bolometric luminosity $L$. This is done using the
analytical expression
$$
{ {\ell_x} \over \ell} = 1.98 + 0.26 \log {\ell} \eqno(5)
$$
which fits the values of $\ell_x$ for numerical models to better than
4 \%.

We account for the presence of a relic magnetic field
assuming that emission is now concentrated in the polar caps.
All radiative effects of the magnetic field are neglected, so its only role
is to channel
the accretion flow. A different approach was considered by Nelson {\it et
al.\/} (1995), who evaluated the contribution of the cyclotron line and
supposed
that the thermal part of the emission is blackbody. For very high fields,
the atmosphere
is no more in LTE and a detailed analysis of radiative transfer
including magnetic effects is required (see e.g. Miller 1992). In addition, in
this case magnetic pressure could temporarily stop the incoming flow, and
the accretion process may be cyclic with recurrence time $\sim 10^6$ s
(Treves, Colpi \& Lipunov 1994).
For this reason, here we limit our discussion to the case of
$B \sim 10^9$ G.
The total emitting area is now $2 A_c$, where $A_c = \pi r_c^2$ is the
area of the polar cap, $r_c = \rs^{3/2} r_A^{-1/2}$ and $r_A$ is
the Alfv\`en radius:
$$
\eqalignno{
r_A & = \left ( { {B^2 \rs^6} \over {\sqrt{2GM} \Mdot}} \right )^{2/7} & \cr
 & = 4.3 \times 10^7 (1 + \chi)^{-2/7}
n_H^{-2/7}\left ({B \over 10^9 G} \right )^{4/7}
\left ({v \over { {\rm km \, s^{-1}}} } \right )^{6/7} \, {\rm cm}\, .& (6)\cr}
$$

The limited size of the radiating region produces a hardening of the spectrum
with respect to the unmagnetized case with the same luminosity,
since now it is $F = L/2A_c > L/4\pi\rs^2$.

\beginsection 3. THE NEUTRON STARS DISTRIBUTION FUNCTION

ONSs accretion luminosity strongly depends on the star velocity with
respect to the ISM and on the local density of the surrounding material,
so, in order to shed light on their collective
observational properties, the knowledge of the present ONSs
distribution function in phase space is needed. Unfortunately,
the velocity distribution of pulsars at birth is still poorly known, and
any evolutionary scenario remains affected by this indetermination.
In a detailed study, Narayan \& Ostriker (1990) found that observational
data of periods and magnetic fields for a sample of about 300
pulsars are well fitted by invoking the presence of two
populations of neutron stars at birth, slow (S) and fast (F) rotators.
These two populations differ in their
kinematical properties, and the F rotators are characterized by a
mean velocity and by a scale height lower than the S ones.
Basing on this picture, the time evolution of the distribution
function in the galactic potential has been studied by several
authors (Blaes \& Rajagopal 1991; BM; see also Paczy\'nski 1990).
In a recent paper, basing on a sample of 29 young pulsars, Lyne \& Lorimer
(1994) suggested the
possibility that neutron stars are born with typical velocities higher
than both the F and S populations of Narayan \& Ostriker.
Their sample, however, is not complete and this result would {\it prima
facie\/} imply that most pulsars evaporate from globular clusters and
the Galactic plane. For this reason we retain the velocity distribution
at birth proposed by Narayan \& Ostriker. In any case, results
presented by BM show that the contribution of high--velocity objects is really
negligible as far as the observability of ONSs is concerned.

Detailed tabulations of the
distribution function are not available in the literature, so
we have repeated the calculations
presented by BM, evolving spatial and velocity distributions corresponding
to the F population of model b of Narayan and Ostriker (1990). All NSs were
assumed to be born at $t=0$.
The Galactic potential is sum of three contributions, disk, spheroid and halo,
and is taken from Blaes \& Rajagopal (1991).
The fraction of F objects is 55 \% of the total number of neutron stars,
and we use $N_{tot} = 10^9$. The distribution function was calculated
integrating the orbits of 48000 stars, up to $t=10^{10}$ ys. Initial
conditions at $t=0$ were obtained assuming that the velocity distribution
is gaussian in each component, together with the vertical distribution;
the radial one is poissonian and the azimuthal one is uniform. Orbits were
accepted only if the final parameters were in the ranges $0<R<20$ kpc,
$|z| < 2.5$ kpc and $|v|< 200$ km/s. We have found more convenient for later
applications (see next sections) to store the distribution function in
heliocentric coordinates, $f=f(r,b,l,v)$. The intervals in heliocentric
coordinates and in $v$ were divided in 75 equally spaced bins.
In order to compare the output of our simulation
with the results of BM, we performed the evolution of
the distribution function for 20000 stars
using galactocentric coordinates: in the local region $7.5 \ {\rm kpc} \leq R
\leq 9.5 \ {\rm kpc}$
we find $\langle z^2 \rangle^{1/2} = 739$ pc;
$\langle z \rangle = 261$ and 249 pc in the northern and southern hemisphere,
respectively. The mean velocity in this local region, averaged
over $|z| \leq 200$ pc, turns out to be 78 km/s, and
$\langle v^2 \rangle^{1/2} = 87$ km/s. All these values are in close agreement
with the results of BM.

In the next sections, we will use $f(r,b,l,v)$ to estimate the
observability of ONSs as X--ray sources.
However, mainly in the case of spherical emission, a detectable
X--ray flux at earth is produced only by the small number of
stars that populate the low--velocity,
small--$r$ tail of the distribution. In this case a reliable statistical
sample could
be obtained only from the evolution of
a large number of stars ( $\sim$1\% of
the total population), requiring an unacceptably high computational time.
In the alternative hypothesis in which the birth--rate of NSs is assumed
to be constant, the statistics could be improved without increasing the number
of computed orbits, as proposed by BM.
In this case, in fact, it is convenient to store the distribution function
every $10^6$ ys: the final distribution can
be obtained as the superposition of different outputs and will be
representative of stars born at different times with the same initial
conditions. In our case, however, we need to know the dependence of $f$ on all
spatial coordinates, and the storage and handling of such a high number, $\sim
10^4$, of very large matrices, $75\times 75\times 75\times 75$, become
troublesome.

To circumvent this problem, we have calculated the
local velocity distribution $g(v)$, averaged over the region $r \leq 2$
kpc,
and a best fit $G_{fit}(v)$ of its
cumulative function $G(v)=\int_0^v g(v')\, dv'$. We found that a function of
the form
$$
G_{fit} (v) = {{\left ( v/v_0 \right )^m} \over { 1 +
{\left ( v/v_0 \right )^n}}} \eqno(7)
$$
\noindent with $ v_0 = 69.0 \, {\rm km \, s}^{-1}$, $n\simeq m = 3.3$,
provides a good fit to $G(v)$.
Figure 2 shows $G(v)$ at $t \, = \, 0$, after $t \, = \, 10^{10} \, {\rm ys}$
and $G_{fit}$. In this local region, the number
density of stars turns out to be $ n_0 = 3\times 10^{-4}(N_{tot}/10^9)
\, {\rm pc^{-3}}$, which is about 2.5 times lower than the midplane
density found by BM. This apparent discrepancy is due to the fact that
we are averaging over a sphere of radius 2 kpc, centered on the sun. By
comparison the star density computed in BM at $z=1$ kpc is about $1\times
10^{-4}(N_{tot}/10^9)\, {\rm pc^{-3}}$.
It can be seen that the relative number of low velocity stars
is higher than the initial one, reflecting the fact that
they do not move significantly away from the Sun during
their evolution, as already noted by BM.
The results of the numerical simulations are also confirmed by an alternative
and faster semi--analytical
approach based on an approximation of the 3th integral of the motion for low
velocity stars (BM).

\beginsection 4. THE CONTRIBUTION TO THE SOFT XRB

The possibility that a new, yet undetected, population of X--ray sources
contributes to the X--ray background, soothing the problem of the
observed soft excess, has been recently suggested by Hasinger {\it et al.\/}
(1993). By carrying out a
detailed analysis of 27 fields at high galactic latitude,
$ |b| \geq 30^{\circ}$, these authors found that, at the faintest
flux limit of ROSAT, about 60 \% of the
background is resolved into extragalactic discrete sources (see also Comastri
{\it et al.\/} 1995). Their flux is $1.48 \times 10^{-8} {\rm erg \, cm^{-2}
s^{-1} sr^{-1}}$ in the 0.5--2 keV band  over a total flux
$F_{XRB}=2.47 \times 10^{-8} {\rm erg \, cm^{-2} s^{-1} sr^{-1}}$ in the same
energy band.
The projected number density of resolved sources turns out to be
$413 \, {\rm deg^{-2}}$, which exceeds by about 60 \%
the density of QSOs predicted by standard evolutionary models.
Even assuming that
at the same flux limit the contribution of stars is $\sim$ 10 \%, about
$120 \, {\rm sources \, deg^{-2}}$ remain to be explained, together with
$\sim 40 \%$ of the XRB (soft excess).
As noted by Hasinger {\it et al.\/}, these results can be interpreted
either by a more complicated model for the evolution of the QSOs
X--ray luminosity function
or invoking the presence of a new population of sources.
An important point emerging from ROSAT observations that sets constraints
on the nature of the new population
is that the average spectrum of resolved sources becomes harder at lower
flux limits.

Recently MG discussed the possibility that the required new population is
galactic in origin and derived its main properties,
assuming that it is composed of standard candles of fixed luminosity
$L$. In particular, they concluded that
objects with a typical X--ray luminosity
$\sim 10^{30}$ erg/s, and with a local density of $\sim
10^{-4}-10^{-5} \ {\rm pc^{-3}}$ could explain
a substantial fraction, 20--40 \%, of the XRB, giving, at the same
time, the required source density of $\sim 120 \, {\rm deg^{-2}}$.
Assuming a blackbody spectrum at the star effective temperature, MG have
ruled out the possibility that
accreting ONSs could be the proposed population, just
on the basis of their spectral properties. The emitted spectrum, in fact,
is too soft, with a mean photon energy $\sim $ 30 eV for
$L\sim 10^{30}-10^{31}$ erg/s. The role of the magnetic field which
drastically diminishes the emitting area, and
the harder spectra obtained from our calculations,
together with their hardening with decreasing luminosity, lead us to
reconsider the possibility that ONSs can contribute to the XRB.

Let us first consider the contribution of the entire
ONSs population to the XRB. Here and in the following, we
use the computed distribution function and the ISM density discussed in
section 2 to evaluate the actual ONSs luminosity.
The presence of the Local Bubble of radius $\sim$ 100 pc
surrounding the Sun has been neglected, since the fraction
of stars that falls within this underdense region turns out
to be negligible. The emitted flux per unit solid angle is:
$$
\left < { { dI} \over {d \Omega}} \right > = {1 \over {4 \pi}}
\int_0^{v_{max}}
\int_{\Omega} \int_0^{r_{max}} f \left (r,b,l,v \right ) { { \tilde L_x}
\over {4 \pi r^2}} r^2 dr d\Omega dv\, \eqno(8) ,
$$
where the integrals in $r$, $\Omega$
span the entire Galaxy and $\tilde L_x$ is the
source luminosity in the 0.5--2 kev band, corrected for interstellar
absorption.
The upper limit in the velocity integral
is set by the fact that no synthetic spectra are available for
$\ell < 4 \times 10^{-8}$ which corresponds to a minimum flux of
$3.8 \times 10^{17}\, {\rm erg \, cm^{-2} s^{-1}}$ at the
source.
This is of little importance, since equation (5)
shows that $\ell_x$ becomes vanishingly
small for $\ell \mincir 2 \times 10^{-8}$.

Next we consider the number of resolved sources which is given by
$$
\left < { { dN} \over {d \Omega}} \right > = {1 \over {4 \pi}}
\int_0^{v_{max}}
\int_{\Omega} \int_0^{d_{max}} f \left (r,b,l,v \right ) r^2 dr d\Omega dv
\eqno(9)
$$
where $d_{max}$ is the maximum distance at which a star of
luminosity $\ell$ produces a count rate above the threshold of Hasinger
{\it et al.\/} (estimated to be $\sim 2\times 10^{-4} \, {\rm counts \,
s}^{-1}$,
corresponding approximately to a flux of $2 \times 10^{-15} {\rm erg\, cm^{-2}
 s^{-1}}$).
Both calculations have been repeated in the case of polar cap accretion
with $B=10^9 $ G; the total column density
corresponding to a given star position was used.
The spatial distribution of ONSs has a characteristic scale height
of $\sim 260$ pc, so both contributions, to the XRB and to the
number count of sources, depend on direction. In order to
quantify the degree of anisotropy, the two calculations have been
repeated for objects with high ($|b| \, \ge \,
30^{\circ}$) and low $(|b| \, \le \, 30^{\circ})$ galactic latitude. The
first case allows a closer comparison with Hasinger {\it et al.\/} and MG.

Our results are summarized in table 1.
In the case of spherical emission the ONSs contribution to the
XRB is really unimportant ($<1$ \%) similarly to the finding of MG \& BM.
Only a small fraction of stars, those in the low velocity tail
of the distribution, have spectra hard enough to
give a non negligible contribution to the background intensity.
Our result is, anyway, $\sim$ 40\% greater than the estimate by BM.

In contrast, when a non--zero magnetic field is taken into
account, the ONSs contribution to the XRB,
averaged over all latitudes, is found to be
of order 10\%, and 5--6 \% at
the high latitudes considered by Hasinger {\it et al.\/} and MG. This
corresponds to 12--25\% of the observed soft excess. As can be seen from
table 1, the ONSs contribution is higher by a factor 2--3 in the galactic
plane, implying a certain degree of anisotropy. However, this is not in
contrast with the observed near--isotropy of the XRB
at energies in the 0.5--1 keV band (see e.g. Mc Cammon \& Sanders 1990)
since ONSs can contribute at most up to 25\% of the soft excess. At the
sensitivity limit of the deep survey of Hasinger {\it et al.\/},
the number density of resolved sources, averaged over all latitudes, is $\sim
20 \ {\rm deg}^{-2}$, which amounts to about 5\% of the total and to $\sim$
18 \% of the non--QSOs, non--stellar component.

{}From a direct analysis of the on line catalogue of the ROSAT PSPC pointings
(ROSATSRC),
we found that the mean number of sources detected with flux larger
than $10^{-3} \, {\rm counts \, s}^{-1}$
($ \sim 10^{-14} \, {\rm erg \, cm^{-2} s^{-1}}
$),
is $\sim 30\, {\rm deg}^{-2}$
($b < 30^\circ$) and $\sim 40\, {\rm deg}^{-2}$ ($b > 30^\circ$).
This result is formally in agreement with the mean number of unidentified
sources predicted by our model at the same flux limit, $\sim 10\,
{\rm deg}^{-2}$ (see the next section), although it is apparent that the ratio
of ONSs over the total number of sources is unacceptably high.
The contribution of ONSs to the soft X--ray background would scale as the
percentage of sources which will be identified as ONSs.

The estimated ONSs contribution to the diffused emission is not probably
strongly affected by the limits of the statistical
analysis and the details of the ISM distribution, as it is shown below.
Assuming a mean value
$n_H = 0.5 \, {\rm cm^{-3}}$, $B = 10^9$ G and using equation (5), the
X--ray luminosity is:
$$
\ell_x =\cases{
\left (3.5 \times 10^{-4} - 5.5 \times 10^{-5} \ln v \right )/v^3,
& $ v \ge c_s$,\cr
\, \cr
2.8 \times 10^{-7}, & $v \le c_s$.\cr}
$$
In order to estimate the contribution provided by the most luminous
stars, we can expand the approximate cumulative function, discussed in
section 3, for $(v/v_0)^n\equiv x^n\ll 1$, to get ($m \simeq n$)
$$
f \left(v \right ) = { {d G_{fit} \left ( v \right )} \over {dv}}
\simeq {m \over {v_0}} x^{m-1} \left ( 1 - 2 x^n \right )\, .
$$
The mean source luminosity is then
$$
\langle \ell_x\rangle\simeq \int_0^{v_{max}} \ell_x \left(v \right ) f \left(v
\right ) dv\, ,
$$
where $v_{max}$ is the largest velocity at which our expansion can be reliably
used. If we assume that the
largest emission comes from stars within a volume $V$ of radius 1 kpc
with density $n_0=3\times
10^{-4} \ {\rm pc}^{-3}$ and at a mean distance $d \simeq 500$ pc,
the contribution to the XRB turns out to be:
$$
{\langle \ell_x \rangle \over {4 \pi d^2}} {n_0 \over {4\pi}}
V = 1.23\times 10^{-9} \ {\rm erg\, cm^2\, s^{-1}\, sr^{-1}}\sim 0.05 F_{XRB}
\, ,
$$
that is of the same order
of that obtained from the numerical simulation.

\beginsection 5. OBSERVABILITY OF INDIVIDUAL SOURCES

In this section we rediscuss the observability of accreting ONSs, following
previous studies on this subject (TC; BM; CCT).
In particular we refer to PSPC on board of ROSAT;
the response curves, needed in equation (4) to obtain the count
rate, are taken from ROSAT guide for observers.
Due to the combined differential effects
introduced
by the response of the detector and the absorption of the ISM, this analysis
crucially relies both on the spectral shape and on the mean photon energy.

As can be seen from
equation (2), typical values for $n_H$ are about $0.2-0.5 \,{\rm
cm^{-3}}$. Higher values seem to occur in the Sco--Cen
and Per directions, where $n_H\sim 1-10\,{\rm
cm^{-3}}$. As far as the observability of individual sources is
concerned, we assume a uniform medium and treat $n_H$ as a free parameter.
This will also allow a closer comparison of our results with those obtained
in previous investigations (TC; Madau \& Blaes 1994).
We consider here three typical
values for the density,
$n_H = 0.2,0.5, 1 \, {\rm cm^{-3}}$,
as representative of different lines of sight.

A star of luminosity $\overline L$ (i.e. moving at $v=\overline v $)
can be observed up to a
maximum distance $d_{max}$ at which
the count rate $N$ (eq.[4])
becomes lower than the sensitivity limit of the
detector ($1.5 \times 10^{-2} \, {\rm counts/s}$ for the all sky survey,
ASS, and
$10^{-3} \, {\rm counts/s}$ for Deep Exposure, DE). Table 2 shows $d_{max}$
for two typical luminosities, $\ell = 10^{-7}$ and $\ell=4 \times 10^{-8}$:
at distances
$d \leq d_{max}$ all stars with $L \geq \overline {L}$ (i.e.
$v \leq \overline {v}$) can be observed. For comparison the corresponding
quantities, as derived for a blackbody spectrum are listed in parenthesis.

The actual
distribution function $f(r,v,l,b)$ of ONSs was derived in section 3;
since the scale height of the distribution,
$\sim 250$ pc, is comparable to that of the gas,
$f$ can reasonably provide an estimate also of
the actual distribution of accreting ONSs, so we could, in principle,
calculate the expected number of observable objects within $ d_{max}$:
$$
N_{ons} \left(\leq d_{max}, \leq \overline{v} \right ) = \int_{\Omega}
\int_0^{d_{max}}
\int_0^{\overline{v}} f r^2 dr d \Omega dv \eqno(10)
$$
where $r$ is the radial distance from the Sun.
To avoid possible fluctuations due to the small number of slow stars
in the local region present in our simulation,
we replaced the previous formula with
$$
N_{ons}\left(\leq d_{max}, \leq \overline{v} \right ) = \cases{
{4 \over 3}\pi d_{max}^{3} n_0 G_{fit} \left ( \overline {v} \right) &
$d_{max}<
z_{ISM}$,\cr
\, \cr
2\pi d_{max}^{2} z_{ISM}n_0 G_{fit} \left ( \overline {v} \right) & $d_{max}>
z_{ISM}$,\cr}
\eqno(11)
$$
where $n_0 \, = 3 \times 10^{-4} \, {\rm pc}^{-3}$.
The typical scale height for the ISM is $z_{ISM}=300$ pc; ONSs with
$z$ larger than $z_{ISM}$ are assumed not to accrete.
The corresponding values are reported in table 2.
Eq. (11) contains, nevertheless, two major simplifications:
first, we assumed that
locally the spatial and velocity dependence of the distribution function
can be factorized; second, we considered
uniformly distributed ONSs. Since this could invalidate our statistical
analysis, we stress that
different columns in the table do not have the same reliability.
A comparison with the corresponding values calculated for a blackbody
shows that maximum distances computed for actual spectral distributions
are systematically higher
and in high--density regions can become greater by a factor 2--3.
All calculations have been repeated for polar cap accretion with
$B = 10^9$ G and for the same values of $\ell$ (see table 3).

We note that stars with $\ell \approx 10^{-8}$ are detectable up to typical
distances $\approx 200-300$ pc for ASS, even if they emit from the entire
surface.
A comparison with TC results shows that these
values are of the order of their estimates for polar cap accretion.
However, we stress that in our case the spectral hardening
is only due to radiative processes; when a non zero magnetic field is
taken into account, objects with the same luminosity become visible
at larger distances. As it is apparent from table 3, if the
magnetic field is $10^9$ G
about 10 sources ${\rm deg^{-2}}$ are expected to be above
the DE ROSAT threshold; this figure is a factor $\sim 10$ greater than
the estimates of TC and CCT.

It has been already stressed (BM; CCT) that giant molecular clouds
in our galaxy provide the most favorable environment for the occurrence of
high accretion rates. We reconsider this possibility, referring to the sample
of 18 nearby clouds studied by Dame {\it et al.\/} (1987).
Assuming that the clouds are spherical and homogeneous, the values of their
radius $R_c$ and density $n_c$ has been calculated by CCT. A
luminosity $\ell \, = \, 10^{-7}$,
which correspond to typical velocities in the
range $40-90 \, {\rm km \, s}^{-1}$ has been assumed.
Since velocities are largely supersonic, the use of equation (1)
is fully justified. Being the local density much higher than the typical ISM
density, we assume
that all absorption is due to the cloud material. Table 4 shows the
count rate (CR) for the calculated spectrum and the expected number
of observable ONSs:
$$
N_{ons} \left(\leq v \right ) =
{4 \over 3}\pi R_{c}^{3} n_0G \left ( v \right); \eqno(12)
$$
the blackbody count rate is also reported for comparison.
The count rate turns out to be above the threshold of Rosat (in DE)
for more than half of the clouds in the spherical case,
due to the hardening af the spectra with respect to a blackbody one.
Clouds 15--16 and 18 appear also promising because of their
high expected count rates, although the expected number of ONSs is small.
We note, however, that, since the number of neutron stars in the clouds are
small, large statistical fluctuations are possible.
The corresponding values for polar cap accretion are summarized in table 5.
Clouds 5--7, 9--13 represents the most favorable sites for observability and
this agrees with the previous estimate presented by CCT.

\beginsection 6. DISCUSSIONS AND CONCLUSIONS

By computing the present distribution function of ONSs
and using a map of neutral and molecular gas in the local region,
we performed an analysis of the ONSs contribution to the diffuse emission in
the soft X--ray energy band. A recent calculation of the spectrum emitted by
accreting neutron stars (Zampieri {\it et al.} 1995) shows that,
for luminosities $\sim 10^{30}-10^{31}\, {\rm erg \,s^{-1}}$,
a relevant fraction (10\%--20\%) of the total flux
is emitted in the 0.5--2 keV band even in the absence of magnetic field.
Furthermore, the
spatial density of ONSs in the solar neighbourhood
computed using the evolved distribution function
turns out to be $n_0 = 3 \times 10^{-4} \left (N_{tot}/10^9 \right )
\, {\rm pc^{-3}}$. Both these values
are in agreement with the constraints set by MG on possible Galactic
candidates which may contribute substantially to the XRB.
We found that, in
the case of polar cap accretion ($B \sim 10^9$ G),
when a greater number of objects
produce spectra hard enough to be detectable,
the ONSs contribution
is $\approx 10 \%$ of the total measured intensity in the 0.5--2 keV band,
corresponding to 25--50\% of the observed soft--excess.
The fraction of resolved
sources at the limit of the deepest ROSAT surveys is about 5 \%. However,
the strongest constraint follows from the expected number of sources above
$10^{-3} \, {\rm counts \, s}^{-1}$ and this will
determine the importance of
the ONSs contribution to the X--ray background.

We note that these results are not strongly dependent on the
overall shape of the velocity distribution of ONSs, but only on the
number of objects in the
low velocity tail, here calculated using the Narayan \& Ostriker
distribution function.
For a high velocity distribution at birth, such as that recently
presented by Lyne \& Lorimer (1994), the extrapolation to low velocities
seems to indicate that the number of stars with velocity less than $\sim 30$
km/s, turns out of the same order. However one should keep in mind
that the distribution in the low velocity tail of ONSs remain very uncertain;
for instance, it can  be affected by
dynamical heating, as suggested by Madau \& Blaes (1994).
This process, observed in the local disk star population,
causes the velocity dispersion to increase with age as
a consequence of scattering by molecular clouds and spiral arms
(Wielen 1977). If ONSs participate the same process, dynamical
heating over the lifetime of the Galaxy may scatter a fraction of
low velocity stars to higher speeds.
This could have a major effect on the source number counts (that
can be decreased  up to a factor 10) and it may reduce the contribution
of luminous ONSs to the background.
Other factors of indetermination, like the poor knowledge about
the NSs birth rate, and hence, of their present total number, may affect our
conclusions to the same extent.
In addition, present estimates of the magnetic field of ONSs are subject
by various uncertainties.

We conclude that,
even if neutron stars do not account completely for the characteristics
of the galactic population proposed by Hasinger et al. (1993) and
MG, their contribution  may be of importance.

\beginsection ACKNOWLEDGEMENTS

We thank the referee, Piero Madau, for some useful and constructive
comments on the
manuscript, Michiel van der Klis for stimulating discussions
and Tomaso Belloni for his help in extracting data
from ROSATSRC.

\vfill\eject

\beginsection REFERENCES

\ref{Blaes, O., \& Rajagopal, M. 1991, ApJ, 381, 210}
\ref{Blaes, O., \& Madau, P. 1993, ApJ, 403, 690 (BM)}
\ref{Bloemen, J.B.G.M. 1987, ApJ, 322, 694}
\ref{Colpi, M., Campana, S., \& Treves, A. 1993, A\&A, 278, 161 (CCT)}
\ref{Comastri, A., Setti, G., Zamorani, G., \& Hasinger, G. 1995, A\&A, in
press}
\ref{Dame, T.M. {\it et al.\/} 1987, ApJ, 322, 706}
\ref{Danner, R., Kulkarni S.R., \& Hasinger G. 1995, in Proceedings of the 17th
Texas Symposium, in press}
\ref{De Boer, H., 1991, in Proceedings of the 144th IAU Symposium, H. Bloemen
ed. (Kluwer: Dordrecht), 333}
\ref{Dickey, J.M., \& Lockman, F.J. 1990, ARA\&A, 28, 215}
\ref{Hasinger, G., {\it et al.\/} 1993, A\&A, 275, 1}
\ref{Helfand, D.J., Chanan, G.A., \& Novick, R. 1980, Nature, 283, 337}
\ref{Lyne, A.G., \& Lorimer, D.R. 1994, Nature, 369, 127}
\ref{Madau, P., \& Blaes, O. 1994, ApJ, 423, 748}
\ref{Maoz, E., \& Grindlay, J.E. 1995, ApJ, in press (MG)}
\ref{Mc Cammon, D., \& Sanders, W.T. 1990, ARA\&A, 28, 657}
\ref{Miller M.C., 1992, MNRAS, 255, 129}
\ref{Morrison, R., \& McCammon, D. 1983, ApJ, 270, 119}
\ref{Narayan, R., \& Ostriker, J.P. 1990, ApJ, 270, 119}
\ref{Nelson, R.W., Wang, J.C.L., Salpeter, E.E., \& Wasserman, I. 1995, ApJ,
in press}
\ref{Novikov, I.D., \& Thorne, K.S. 1973, in Black Holes, C. DeWitt and
B.S. DeWitt eds. (Gordon \& Breach: New York), 343}
\ref{Ostriker, J.P., Rees, M.J., \& Silk, J. 1970, Astrophys. Letters, 6,
179}
\ref{Paczy\'nski, B. 1990, ApJ, 348, 485}
\ref{Pounds, K.A., {\it et al.\/} 1993, MNRAS, 260, 77}
\ref{Stocke, J.T., {\it et al.\/} 1995, AJ, in the press}
\ref{Treves, A., \& Colpi, M. 1991, A\&A, 241, 107 (TC)}
\ref{Treves, A., Colpi, M., \& Lipunov, V.M. 1993, A\&A, 269, 319}
\ref{Treves, A., Colpi, M., Turolla, R., Zampieri, L., \& Zane, S. 1995,
to appear in Proceedings of the 7th Grossmann Meeting}
\ref{Wielen, R. 1977, A\&A, 60, 263}
\ref{Zampieri, L., Turolla R., Zane S., \& Treves, A. 1995, ApJ, 439, 849}
\vfill\eject

\beginsection FIGURE CAPTIONS

\refig{Figure 1.\quad Emergent spectra for $L =
10^{-7}$ and $4 \times 10^{-8} \, L_{Edd}$ (full lines), together with the
corresponding blackbody spectra
at the neutron star effective temperature (dashed
lines).}
\medskip

\refig{Figure 2.\quad $G(v)$ at $t \, = \, 0$ (dashed line) and
after $t \, = \, 10^{10} \, {\rm ys}$ (continuous line);
$G_{fit}$ (dash--dotted line) is also shown.}

\bye

%
\nopagenumbers
\magnification=1200
\tolerance=1000
\null
\vskip 1.5truecm
\centerline{Table 1}\medskip
$$\vbox{\tabskip=1em plus2em minus.5em
\halign to\hsize{#\hfil &\hfil # \hfil & \hfil # \hfil &
\hfil # \hfil &
 \hfil # \hfil  & \hfil # \hfil & \hfil # \hfil &
\hfil # \hfil & \hfil # \hfil &\hfil # \hfil &\hfil #
\hfil\cr
&$\,$  &$\,$  &$\displaystyle{ \left < { {dI} \over {d \Omega}} \right >}^
{\rm{a}} $
 &
$\displaystyle{\left < { {dN} \over {d \Omega}} \right >}^{\rm {b}}  $ &
$ \displaystyle{{{\left <  dI/d \Omega \right > }\over {F_{XRB}}}
\times 10^2} $ &
$ \displaystyle{{ {\left < dN/ d \Omega \right >} \over 413} \times 10^2} $
& & & & \cr
\noalign{\bigskip} \cr
%
%
%
%
&  unmagnetized  & $\,$ & $6.3 \times 10^{-11}$  &   0 & $2.5 \times 10^{-3}$
&  0  & & & & \cr
\noalign{\medskip} \cr
%
& $B = 10^9   $ G  & all $b$ & $2.4 \times 10^{-9} $
&  22 & $9.8 $
& $5.3  $ & & & & \cr
& & & & & & & & & & \cr
& $\,$ & $\vert b \vert \, \ge \, 30^{\circ} $ & $1.3 \times 10^{-9} $
&  6 & $5.2 $
& $1.4$  & & & & \cr
& & & & & & & & & & \cr
& $\,$ & $\vert b \vert \, \le \, 30^{\circ} $ & $3.6 \times 10^{-9} $
&  38 & $14.5 $
& $9.2  $ & & & & \cr
& & & & & & & & & & \cr
%
%
%
 }}$$

\bigskip
\parindent 0.truept
$^{\rm a}$ ${ \rm erg \, cm^{-2} s^{-1} sr^{-1}} $

\parindent 0.truept
$^{\rm b}$ $ {\rm sources \, deg^{-2} } $

\bye

%

\magnification=1200
\tolerance=1000

\null
\vskip 1.5truecm
\hsize=25.truecm\vsize=6.5in
\nopagenumbers
\baselineskip=24truept
\def\rho{\varrho}
\def\Mdot{\dot M}
\def\mincir{\raise -2.truept\hbox{\rlap{\hbox{$\sim$}}\raise5.truept
\hbox{$<$}\ }}
\def\magcir{\raise -2.truept\hbox{\rlap{\hbox{$\sim$}}\raise5.truept
\hbox{$>$}\ }}
\centerline{Table 2}\medskip
$$\vbox{\tabskip=1em plus2em minus.5em
\halign to\hsize{#\hfil &\hfil # \hfil & \hfil # \hfil & \hfil # \hfil &
 \hfil # \hfil  & \hfil # \hfil & \hfil # \hfil &
\hfil # \hfil & \hfil # \hfil &\hfil # \hfil &\hfil # \hfil\cr
& $\displaystyle{{{L} \over {L_{Edd}}}} $ & $n_H \left ( {\rm cm}^{-3} \right )
$
& $\overline{v} \left ( {\rm km \  s}^{-1} \right )$
& $d_{max}^{a} \left( {\rm pc} \right )$
&$N_{ons} \left(\leq d_{max},\leq \overline{v} \right)^{a}$ &
$d_{max}^{b}\left ( {\rm pc} \right )$
&$N_{ons} \left(\leq d_{max},\leq \overline{v} \right)^{b}$&
 & & \cr
\noalign{\smallskip}
%
%
\noalign{\bigskip\medskip}
& $         10^{-7}$ & 0.2 &  9.9 & $ 525 \  (375) $ & $ 260 \  (132)$ &
$1170 \  (725)$ & $1293 \  (497)$ & & &  \cr
& $         10^{-7}$ & 0.5 & 13. & $ 360 \  (240) $ & $333 \  (148)$ &
$845 \  (425)$ & $1837 \  (465)$ & & &  \cr
& $         10^{-7}$ & 1.  & 17. & $ 275 \  (165) $ & $253 \  (55)$ &
$675 \  (275)$ & $2495 \  (414)$ & & &  \cr
& $4 \times 10^{-8}$ & 0.2 & 14. & $ 380 \  (220) $ & $402 \  (135)$ &
$840 \  (445)$ & $1965 \  (552)$ & & &  \cr
& $4 \times 10^{-8}$ & 0.5 & 19. & $ 265 \  (145) $ & $312 \  (51)$ &
$555 \  (275)$ & $2324 \  (570)$ & & &  \cr
& $4 \times 10^{-8}$ & 1.  & 23. & $ 195 \  (105) $ & $262 \  (41)$ &
$415 \  (185)$ & $2737 \  (544)$ & & &  \cr
 }}$$
%
%
%

\bigskip
\parindent 0.truept
$^{\rm a}$ All Sky Survey.

\parindent 0.truept
$^{\rm b}$ Deep Exposure.


\parindent 0.truept
$z_{{\rm ISM}} = 300 $ pc

\vfill\eject
%

\magnification=1200
\tolerance=1000

\null
\vskip 1.5truecm
\hsize=25.truecm\vsize=6.5in
\nopagenumbers
\baselineskip=24truept
\def\rho{\varrho}
\def\Mdot{\dot M}
\def\mincir{\raise -2.truept\hbox{\rlap{\hbox{$\sim$}}\raise5.truept
\hbox{$<$}\ }}
\def\magcir{\raise -2.truept\hbox{\rlap{\hbox{$\sim$}}\raise5.truept
\hbox{$>$}\ }}
\centerline{Table 3}\medskip
$$\vbox{\tabskip=1em plus2em minus.5em
\halign to\hsize{#\hfil &\hfil # \hfil & \hfil # \hfil & \hfil # \hfil &
 \hfil # \hfil  & \hfil # \hfil & \hfil # \hfil &
\hfil # \hfil & \hfil # \hfil &\hfil # \hfil &\hfil # \hfil\cr
& $\displaystyle{{{L} \over {L_{Edd}}}} $ & $n_H \left ( {\rm cm}^{-3} \right
)$
& $\overline{v} \left ( {\rm km \  s}^{-1} \right )$
& $d_{max}^{a} \left( {\rm pc} \right )$
&$N_{ons} \left(\leq d_{max},\leq \overline{v} \right)^{a}$ &
$d_{max}^{b}\left ( {\rm pc} \right )$
&$N_{ons} \left(\leq d_{max},\leq \overline{v} \right)^{b}$&
 & & \cr
\noalign{\smallskip}
%
%
\noalign{\bigskip\medskip}
& $ 10^{-7}$ & 0.2 &  9.5 & $ 760 \  (705) $ &
$ 480 \  (413)$ & $2480 \  (2110)$ & $5112 \  (3700)$ & & &  \cr
& $ 10^{-7}$ & 0.5 &  13. & $ 685 \  (595) $ &
$ 1069 \  (802)$ & $2055 \  (1675)$  & $ 9563 \  (6353)$ & & &  \cr
& $ 10^{-7}$ & 1.  &  16. & $ 615 \  (515) $ &
$ 1823 \  (1279)$  & $1690 \  (1330)$  & $ 13769 \  (8528)$ & & &  \cr
& $3.  \times 10^{-8}$ & 0.2 & 15. & $ 410 \  (385) $ &
$ 611 \  (539)$ & $1350 \  (1060)$  & $ 6630 \  (4088)$ & & &  \cr
& $3.  \times 10^{-8}$ & 0.5 & 20. & $ 370 \  (305) $ &
$ 1345 \  (914)$ & $1160 \  (845)$  & $ 13224 \  (7017)$ & & &  \cr
& $3.  \times 10^{-8}$ & 1.  & 25. & $ 340 \  (260) $ &
$ 2382 \  (1393)$ & $990 \  (695)$  & $ 20128 \  (9954)$ & & &  \cr
 }}$$
%
%
%

\bigskip
\parindent 0.truept
$^{\rm a}$ All Sky Survey.

\parindent 0.truept
$^{\rm b}$ Deep Exposure.


\parindent 0.truept
$B = 10^{9}$ G

\parindent 0.truept
$z_{{\rm ISM}} = 300 $ pc

\bye
%

\magnification=1200
\tolerance=1000

\null
\vskip 1.5truecm
\hsize=25.truecm\vsize=6.5in
\nopagenumbers
\baselineskip=14truept
\def\mincir{\raise -2.truept\hbox{\rlap{\hbox{$\sim$}}\raise5.truept
\hbox{$<$}\ }}
\def\magcir{\raise -2.truept\hbox{\rlap{\hbox{$\sim$}}\raise5.truept
\hbox{$>$}\ }}
\centerline{Table 4}\medskip
$$\vbox{\tabskip=1em plus2em minus.5em
\halign to\hsize{#\hfil &\hfil # \hfil & \hfil # \hfil & \hfil # \hfil &
 \hfil # \hfil  & \hfil # \hfil & \hfil # \hfil &
\hfil # \hfil & \hfil # \hfil &\hfil # \hfil &\hfil # \hfil &\hfil # \hfil
&\hfil # \hfil\cr
&$\,$ & $ {\rm cloud} $ & $d_c $ & $ R_c $ & $n_c$   &
$CR$ & $CR_{BB}^{\ast}$
&${v}^{a} $ &
$N_{ons}^{a} \left(
\leq v \right ) $& & \cr
&$\,$ & $\,$ & pc & pc & ${\rm cm}^{-3}$   &
${\rm s}^{-1}$ & ${\rm s}^{-1}$
&${\rm km\, s}^{-1}$ &
$ \, $& & \cr
\noalign{\smallskip}
%
%
\noalign{\bigskip\medskip}
&  1 &   Cloud A  & 500 & 20 &  50 & $1.6\times 10^{-3}$
& $2.1\times 10^{-6}$ & 56 & 4 & & \cr
&  2 &   Cloud B  & 300 & 20 &  51 & $4.4\times 10^{-3}$
& $5.2\times 10^{-6}$ & 56 & 4 & & \cr
&  3 &   Cloud C  & 500 & 16 &  67 & $1.5  \times 10^{-3}$
& $1.5\times 10^{-6}$ & 62 &  2 & & \cr
&  4 &   Vul Rft  & 400 & 23 &  61 & $1.7\times 10^{-3}$
& $8.0\times 10^{-7}$ & 60 &  5 & & \cr
&  5 &   Cyg Rft  & 700 & 67 &  29 & $3.3  \times 10^{-4}$
& $8.2\times 10^{-8}$ & 47 & 95 & & \cr
&  6 &   Cyg OB7  & 800 & 64 &  29 & $2.7\times 10^{-4}$
& $7.4\times 10^{-8}$ & 47 & 83 & & \cr
&  7 &   Cepheus  & 450 & 45 &  20 & $2.3\times 10^{-3}$
& $4.4\times 10^{-6}$ & 41 & 18 & & \cr
&  8 &   Taurus   & 140 & 13 & 134 & $9.9\times 10^{-3}$
& $3.1\times 10^{-6}$ & 78 & 1 & & \cr
&  9 &   Mon OB1  & 800 & 34 &  40 & $4.4\times 10^{-4}$
& $2.2\times 10^{-7}$ & 52 &  15 & & \cr
& 10 &  Orion A   & 500 & 27 &  84 & $5.0\times 10^{-4}$
& $8.5  \times 10^{-8}$ & 67 &  12 & & \cr
& 11 &   Orion B  & 500 & 31 &  56 & $7.8\times 10^{-4}$
& $2.4\times 10^{-7}$ & 58 &  13 & & \cr
& 12 &   Mon R2   & 830 & 32 &  36 & $5.0\times 10^{-4}$
& $3.9\times 10^{-7}$ & 50 &  10 & & \cr
& 13 & Vela Sheet & 425 & 26 &  46 & $1.8\times 10^{-3}$
& $1.3\times 10^{-6}$ & 55 &   7 & & \cr
& 14 &    Cham    & 215 & 13 &  44 & $1.5\times 10^{-2}$
& $2.4 \times 10^{-4}$ & 54 &  1 & & \cr
& 15 &  Coalsack  & 175 & 8.5&  65 & $2.4 \times 10^{-2}$
& $4.3\times 10^{-4}$ & 61 &  0 & & \cr
& 16 &   G317--4  & 170 &  5 & 203 & $1.4\times 10^{-2}$
& $1.7\times 10^{-5}$ & 90 & 0 & & \cr
& 17 &    Lupus   & 170 & 18 &  53 & $1.5\times 10^{-2}$
& $2.3\times 10^{-5}$ & 57 &  3 & & \cr
& 18 &    Rcr A   & 150 & 7.6&  68 & $3.5\times 10^{-2}$
& $8.1\times 10^{-4}$ & 62 & 0 & & \cr
 }}$$
%
%
%

\bigskip
\parindent 0.truept
$ ^{a} \, l = 10^{-7}$

\parindent 0.truept
$ ^{\ast} \, T_{eff} = 3.\times 10^{-2}$ kev

\vfill\eject
%

\magnification=1200
\tolerance=1000

\null
\vskip 1.5truecm
\hsize=25.truecm\vsize=6.5in
\nopagenumbers
\baselineskip=14truept
\def\mincir{\raise -2.truept\hbox{\rlap{\hbox{$\sim$}}\raise5.truept
\hbox{$<$}\ }}
\def\magcir{\raise -2.truept\hbox{\rlap{\hbox{$\sim$}}\raise5.truept
\hbox{$>$}\ }}
\centerline{Table 5}\medskip
$$\vbox{\tabskip=1em plus2em minus.5em
\halign to\hsize{#\hfil &\hfil # \hfil & \hfil # \hfil & \hfil # \hfil &
 \hfil # \hfil  & \hfil # \hfil & \hfil # \hfil &
\hfil # \hfil & \hfil # \hfil &\hfil # \hfil &\hfil # \hfil &\hfil # \hfil
&\hfil # \hfil\cr
&$\,$ & $ {\rm cloud} $ & $d_c $ & $ R_c $ & $n_c$   &
$CR$ & $CR_{BB}^{\ast}$
&${v}^{b} $ &
$N_{ons}^{b} \left(
\leq v \right ) $& & \cr
&$\,$ & $\,$ & pc & pc & ${\rm cm}^{-3}$   &
${\rm s}^{-1}$ & ${\rm s}^{-1}$
&${\rm km\, s}^{-1}$ &
$ \, $& & \cr
\noalign{\smallskip}
%
%
\noalign{\bigskip\medskip}
&  1 &   Cloud A  & 500 & 20 &  50 & $2.1\times 10^{-2} $
& $1.2\times 10^{-2}$ & 54 & 3 & & \cr
&  2 &   Cloud B  & 300 & 20 &  51 & $5.6\times 10^{-2} $
& $3.4\times 10^{-2}$ & 54 & 3 & & \cr
&  3 &   Cloud C  & 500 & 16 &  67 & $2.0\times 10^{-2} $
& $1.2\times 10^{-2}$ & 59 &  2 & & \cr
&  4 &   Vul Rft  & 400 & 23 &  61 & $2.6\times 10^{-2} $
& $1.4\times 10^{-2}$ & 58 &  5 & & \cr
&  5 &   Cyg Rft  & 700 & 67 &  29 & $6.7\times 10^{-3}$
& $3.3\times 10^{-3}$ & 45 &  78 & & \cr
&  6 &   Cyg OB7  & 800 & 64 &  29 & $5.3\times 10^{-3}$
& $2.7\times 10^{-3}$ & 45 &  68 & & \cr
&  7 &   Cepheus  & 450 & 45 &  20 & $2.7\times 10^{-2} $
& $1.7\times 10^{-2}$ & 40 &  18 & & \cr
&  8 &   Taurus   & 140 & 13 & 134 & $18\times 10^{-2}              $
& $9.4\times 10^{-2}$ & 75 &  1 & & \cr
&  9 &   Mon OB1  & 800 & 34 &  40 & $6.7\times 10^{-3}$
& $3.7\times 10^{-3}$ & 50 & 12 & & \cr
& 10 &  Orion A   & 500 & 27 &  84 & $1.2\times 10^{-2}$
& $5.4\times 10^{-3}$ & 64 & 10 & & \cr
& 11 &   Orion B  & 500 & 31 &  56 & $1.4\times 10^{-2} $
& $7.4\times 10^{-3}$ & 56 & 13 & & \cr
& 12 &   Mon R2   & 830 & 32 &  36 & $6.9\times 10^{-3}$
& $4.0\times 10^{-3}$ & 48 & 10 & & \cr
& 13 & Vela Sheet & 425 & 26 &  46 & $2.6\times 10^{-2} $
& $1.5\times 10^{-2}$ & 52 & 7 & & \cr
& 14 &    Cham    & 215 & 13 &  44 & $14\times 10^{-2}   $
& $9.6 \times 10^{-2} $ & 52 &  1 & & \cr
& 15 &  Coalsack  & 175 & 8.5&  65 & $22\times 10^{-2}         $
& $15\times 10^{-2}           $ & 59 &  0 & & \cr
& 16 &   G317--4  & 170 &  5 & 203 & $18\times 10^{-2}        $
& $11\times 10^{-2}   $ & 86 & 0 & & \cr
& 17 &    Lupus   & 170 & 18 &  53 & $18 \times 10^{-2}        $
& $11\times 10^{-2} $ & 55 &  2 & & \cr
& 18 &    Rcr A   & 150 & 7.6&  68 & $30\times 10^{-2}          $
& $21\times 10^{-2}               $ & 60 & 0 & & \cr
 }}$$
%
%
%

\parindent 0.truept
$ ^{b} \, B = 10^9$ G, $l = 10^{-7}$

\parindent 0.truept
$ ^{\ast} \, T_{eff} = 1.6 \times 10^{-2}$ kev

\vfill\eject
\bye